\documentclass[pdflatex]{sn-jnl}
\usepackage{graphicx}%
\usepackage{multirow}%
\usepackage{amsmath,amssymb,amsfonts}%
\usepackage{amsthm}%
\usepackage{mathrsfs}%
\usepackage[title]{appendix}%
\usepackage{xcolor}%
\usepackage{textcomp}%
\usepackage{manyfoot}%
\usepackage{booktabs}%
\usepackage{algorithm}%
\usepackage{algorithmicx}%
\usepackage{algpseudocode}%
\usepackage{listings}%
\usepackage[separate-uncertainty=true,list-units=repeat,multi-part-units=brackets]{siunitx}
\usepackage{natbib}
\usepackage{multirow}
\theoremstyle{thmstyleone}%
\theoremstyle{thmstyletwo}%
\theoremstyle{thmstylethree}%
\begin{document}

\title[Morphologic Evolution in Simulated Wood Densification]{Morphologic Evolution in Simulated Wood Densification}

\author*[1]{\fnm{Alessia} \sur{Ferrara}}\email{aferrara@ethz.ch}
\author[1]{\fnm{Júlio O.} \sur{Amando de Barros}}\email{jortiz@student.ethz.ch}
\author[1]{\fnm{Sophie Marie} \sur{Koch}}\email{sokoch@ethz.ch}
%https://orcid.org/0000-0002-1718-4999
\author[1]{\fnm{Falk K.} \sur{Wittel}}\email{fwittel@ethz.ch}
\affil*[1]{\orgdiv{Institute for Building Materials}, \orgname{ETH Zurich}, \orgaddress{\street{Laura-Hezner-Weg 7}, \city{Zurich}, \postcode{8093}, \country{Switzerland}} ORCID: 0009-0003-9627-5200}

\abstract{This study investigates the radial densification of spruce wood using explicit Finite Element Method simulations, focusing on the effects of various densification protocols. These protocols include quasi-static compression, oscillatory excitation, and self-densification through shrinking hydrogel fillings and their impact on the morphogenesis of folding patterns across different tissue types. The simulations incorporate the an isotropic mechanical behavior of wood tracheid walls and account for moisture and delignification effects using a hierarchical approach. Our results reveal the technological potential of targeted densification in creating tailored density profiles that enhance stiffness and strength. These insights offer valuable guidance for optimizing densification processes in practical applications.}


\maketitle

\section{Introduction}\label{sec1}
The densification or compaction of cellular tissue is a practicable solution to extend the use of wood far beyond its native mechanical application limitations. Examples range back centuries from shuttles in weaving looms to shipbuilding, weaponry, bearings, connectors, and supports to aircraft propellers or tonewood - all making use of increased density, stiffness, strength, wear resistance, or related properties \citep{luan_wood_2022,paul_production_2024,sandberg_wood_2017,cabral_densification_2022}. Compaction of wood is mainly achieved in radial direction by simultaneously applying pressure and sometimes heat beyond the glass transition temperature of lignin, to allow for non-destructive folding of the cellular scaffold. Recently, research on delignified softwood, resulting in cellulose scaffolds for chemical modification, fostered new possibilities for wood applications - also in combination with densification processes \citep{frey_delignified_2018,keplinger_etal_2021,song_processing_2018,chaoji_chen_structure-property-function_2020}. It also opened up new technological perspectives to manipulate folding patterns during densification and, consequently, the performance of the densified wood components, which are the core motivation for this numerical work.

The superior performance of densified wood stimulates the desire to simulate the compaction on the tissue scale to understand the influence of process control parameters on the process of densification and final compaction morphologies. Unfortunately, densification simulations are challenging, as they are dominated by geometric non-linearity due to progressive cell wall buckling and multi-surface contact interactions for a disordered cellular structure with highly anisotropic cell walls. Two different simulation approaches have proven to be feasible: the Material Point Method (MPM) and Finite Element Methods (FEM). The application of the MPM for transverse wood densification was first demonstrated by \citet{nairn_numerical_2007} with image-based micro-structure generation \citep{nairn_material_2007}, later extended by Green-elastic material behavior \citep{aimene_simulation_2015}. A refined approach was demonstrated by \citet{perre_new_2016}, confirming the suitability of the method to cope with large geometric non-linearity and multi-body contact. However, simulations are computationally expensive, limiting model sizes, numerical resolution, and simulation times. FEM models on the cellular scale can reach system sizes of entire growth rings and thus become representative \citep{mora_etal_2019}. However, simulations for transverse compaction require dynamic simulations. Only recently, \citet{yan_numerical_2022} demonstrated explicit FEM compaction simulations for a small model with 12 cells with elastic-plastic material behavior. Collaborative folding processes, however require a much larger number of interacting cells. Material non-linearity is essential for the post-densification behavior but not so much for the morphogenesis on the tissue scale, which is the focus of this work. 

Starting from natural tissue sections of spruce wood, we study the evolution of the morphology of cellular folding patterns in computer simulations. We aim to numerically find control parameters for different densification protocols, namely quasi-static mechanical densification, quasi-static mechanical densification with superimposed transverse oscillatory excitation, and shrinkage-induced self-densification through hydrogel filling. For this purpose, we apply explicit Finite Element Method (FEM) simulations with the ABAQUS software on cellular models, whose constitutive properties are calculated on the fly via a hierarchical multi-scale approach and account for moisture and delignification effects. Our work shows that the chosen technological path for densification itself is of great importance for tailoring the resulting material behavior.

%%%%%%%%%%%%%% --- MATERIALS AND METHODS --- %%%%%%%%%%%%%%%%%%%%%%%%%%%%%%%%%%%%%%%%%%%%%%%%%%%%%%%%%%%%%%%%%%%%%%%%%
\section{Materials and Methods}
This section outlines the numerical framework used in this study, beginning with the computation of material properties through a hierarchical multi-scale approach based on composite mixing rules and laminate theory (Sec.~\ref{sec:cellWall}). We then describe the model generation based on processing images of real cellular structures (Sec.~\ref{sec:tissueGen}), to which three densification protocols are applied for simulation (Sec.~\ref{sec:mat_num}).

%%%%%%%%%%%%%%%%%%%%%%%%%%%%%%%%%%%%%%%%%%%%%%%%%%%%%%%%%%%%%%%%%%%%%%%%%%%%%%%%%%%%%%%%%%%%%%%%%%%%%%%%%%%%%%%%%%%%%%
\subsection{Micro-mechanical model for cell wall stiffness}\label{sec:cellWall}
Tracheid cell walls are built up in layers, starting from the surface with the primary (P) and secondary (S) walls that can be considered as a laminate consisting of S1, S2, and S3 sub-layers \citep{fengel_stoll-73,holmberg_persson_1999,persson_micromechanical_2000}. Tracheids are connected via a middle lamella (ML). The ML contains only small amounts of embedded cellulose and presents a 3D random distribution of hemicellulose chains, lignin, and pectin, so rotational symmetry around all axes, resulting in isotropic behavior.

One can discretize cell walls layer-by-layer or apply an equivalent single-layer (ESL) approach with smeared material properties for the composite structures. When layers are numerous or very thin, the layer-wise FEM discretization results in an excessive number of elements that prohibits the simulation of multiple tracheids. In this work, a hybrid approach is followed, where secondary sub-layers (S1, S2, S3) with strong orthotropy form one ESL, while the primary layer (P) and middle lamella (ML) are combined to form an isotropic composite middle lamella (CML). This comes with the disadvantage that layer properties for each cell wall, namely the orthotropic stiffness tensors, must be calculated individually for each cell wall section when defining the model. For this purpose, a hierarchical framework based on composite mixing rules and laminate theory is used, as described in \cite{mora_etal_2019}, initially motivated by the works of \citet{persson_micromechanical_2000,holmberg_persson_1999}.
\begin{figure*}[tb]
  \centering{\includegraphics[width=1\textwidth]{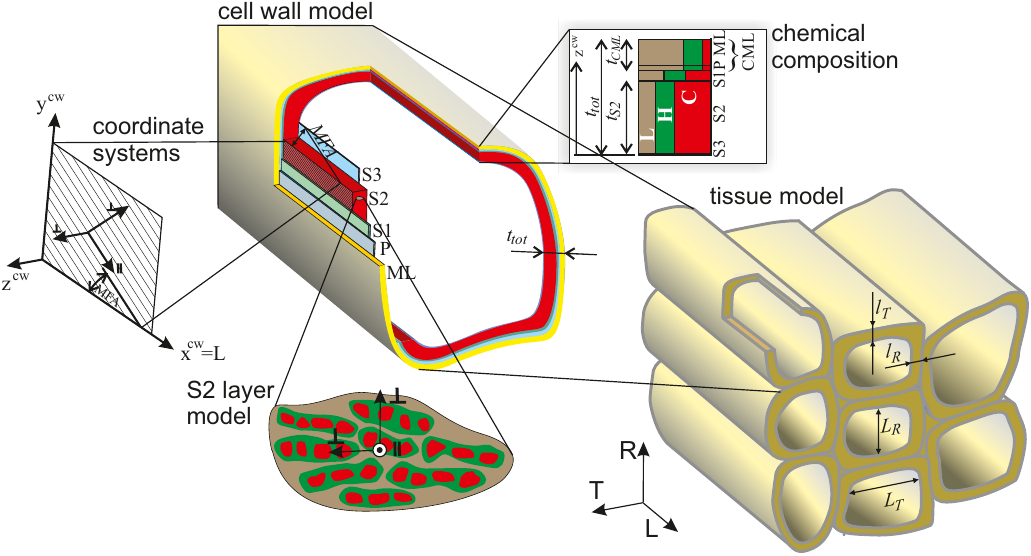}
  \caption{\label{fig:1new} Tissue model with anatomical orientations (L=longitudinal, R=radial, T=tangential), cell-wall model with middle lamellae (ML), primary (P) secondary cell walls (S1, S2, S3) and microfibril angle MFA and cell wall coordinate system ($x^{cw},y^{cw},z^{cw}$), S2 layer model with microfibril aggregates, as well as chemical composition through the cell wall (L=lignin, H=hemicellulose, C=cellulose) and S2 cell wall layer model with the transverse=$\perp$ and MFA=longitudinal=$\parallel$ layer coordinate directions.}}
\end{figure*}

 The P wall exhibits a 2D random orientation of cellulose microfibrils represented by rotational symmetry around one axis and transverse isotropic behavior. As the P wall is very thin compared to the ML and for reasons of numerical stability and simplicity, the P wall and the ML is combined into a compound middle lamella (CML) with a single material that can be interpreted as the matrix that embeds the tracheids (shown in Fig.~\ref{fig:1new}). We assume isotropic behavior for the CML (\num{2} independent material parameters), even though it would exhibit a slight orthotropy in reality. Note that the discrepancies that arise from this assumption were quantified and found to be insignificant in the context of this work. 
 
 For the tracheids, we assume that each S-layer follows material symmetries of a transverse isotropic material within its layer coordinate system (\num{5} independent material parameters). Each layer has an individual cellulose microfibril angle (MFA), as well as a characteristic chemical composition (see Fig.~\ref{fig:1new}, Tab.~\ref{tab:1}). Note that the S1 layer is decomposed into two layers with positive and negative MFA. Interestingly, the respective thicknesses $t_{CML},t_{S1},t_{S3}$ hardly varies across a growth ring (see Table~\ref{tab:1}). Consequently, the thickness of the S2 layer has to vary as $t_{S2}=l_{R,T}-\sum t_i,~\text{with}~(i=CML,S1,S3)$ with $l_{R,T}$ denoting the cell wall thickness in the radial (R) or tangential (R) direction. As we work from image data, we assume each tracheid has a uniform wall thickness $t_{tot}$, from which $t_{S2}$ is calculated. Then, the outer perimeter of each tracheid is down-scaled by $t_{CML}$, so the matrix can fill the space between tracheids (see Sec.~\ref{sec:tissueGen}). Consequently, a stiffness tensor for the matrix, along with an individual stiffness tensor for each tracheid needs to be computed for the FEM model during the pre-processing. This computation starts from the chemical constituents' elastic properties and volume fractions.

For each layer and chemical constituent, the symmetric elastic stiffness tensor $\mathbf{C}$ is of the form
\begin{equation}
\mathbf{C}=\begin{bmatrix}
C_{11} & C_{12} & C_{12} & 0 & 0 & 0\\ 
  & C_{22} & C_{23} & 0 & 0 & 0\\ 
  &   & C_{33} & 0 & 0 & 0\\ 
  &   &   & C_{44} & 0 & 0\\ 
  &  sym &   &   & C_{55} & 0\\ 
 &   &   &   &   & C_{66}
\end{bmatrix},
\end{equation}
where $C_{44}$, $C_{55}$, and $C_{66}$ represent the shear moduli in the planes normal to 1-, 2-, and 3-direction, respectively. While 1-, 2-, and 3-directions are equivalent in isotropic materials, the 1-direction is oriented in microfibril direction ($\parallel$ in Fig.~\ref{fig:1new}) for the transverse isotropic material coordinate systems of the layers. For compactness, we employ vector notation $\mathbf{C}=[C_{11},$ $C_{22},$ $C_{33},$ $C_{44},$ $C_{55},$ $C_{66},$ $C_{12},$ $C_{13},$ $C_{23}]$. Following \citet{cave:1978:1} and \citet{cousins:1976,cousins:1978}, for the chemical constituents cellulose (C), hemicelluloses (H), and lignin (L), the stiffnesses are in GPa:
\begin{align}
\label{eq:7}
\mathbf{C}^C&=\left[137,27,27,6.6,6.6,6.6,14,14,14\right],\\
\mathbf{C}^H&=\left[8,4,4,1,1,1,2,2,2\right],\\
\mathbf{C}^L&=\left[4,4,4,1,1,1,2,2,2\right].
\end{align} 
However, lignin and hemicelluloses are moisture dependent, what is considered by moisture dependent scaling factors, $\alpha^L(u_b^L)$ and $\alpha^H(u_b^H)$, reducing the stiffness tensors $\mathbf{C}^L(u_b^L)=\alpha^L(u_b^L)\cdot\mathbf{C}^L$ and $\mathbf{C}^H(u_b^L)=\alpha^L(u_b^L)\cdot\mathbf{C}^H$, as in \citet{cave:1978:1}. The scaling factors are calculated as
\begin{equation}
\label{eq:8}
\alpha^H\left ( u_b^H  \right ) = h\frac{0.1+0.9\left ( 1-u_b^H \right )}{1+a}, \qquad 
\alpha^L\left (u_b^L  \right )=\frac{1+2\left ( 1-u_b^L \right )}{1+a},
\end{equation}
where $h=6$ and $a=1$ \citep{cave:1978:1}. Each scaling factor $\alpha^i$ is a function of the respective portion of water-responsive sites $u_b^i$, which is a fraction of the bound water $\omega_b$ ($u_b^L=1/3.6\cdot\omega_b$ and $u_b^H=2.6/3.6\cdot\omega_b$). The bound water $\omega_b$ is calculated from the moisture content $\omega$, the bound water at fiber saturation $b=0.2$ ($\omega_{FSP}=0.3$), and the constant $k=0.9$, as
\begin{equation} \label{eq:11}
\omega_b=b\cdot\frac{\left(1-exp\left(-k\cdot\omega/b\right)\right)}{\left(1-exp\left(-0.3\cdot k/b\right)\right)}.
\end{equation}

\begin{table}[htb]
\centering
\caption{Ultra-structural features and chemical composition of cell wall layers with volume contents of cellulose (C), hemicelluloses (H), and lignin (L) following \citet{persson_micromechanical_2000} and \citet{qing_3d_2009}.}\label{tab:1}  
\begin{tabular}{l c c c c c}
\hline
  & \textbf{MFA} & \textbf{thickness} &\multicolumn{3}{c}{\textbf{volume fraction [\SI{}{\percent}]}}\\
& & $\left(\mu m \right)$ & C & \hspace{1.5em}H & L\\ 
\hline\hline
CML & random & 0.35& 12 & \hspace{1.5em}26 & 62 \\ 
%P  & 2D random &  0.175& 12 & 26 & 62 \\
S1 & 60/-60 & 0.125/0.125 & 35 & \hspace{1.5em}30 & 35  \\
S2 & 15 & calculated & 50 &  \hspace{1.5em}27 & 23 \\
S3 & 75 & 0.035 & 45 & \hspace{1.5em}35 & 20 \\
\hline
\end{tabular}
\end{table}

\begin{figure}[htb]
  \centering{\includegraphics[width=1\textwidth]{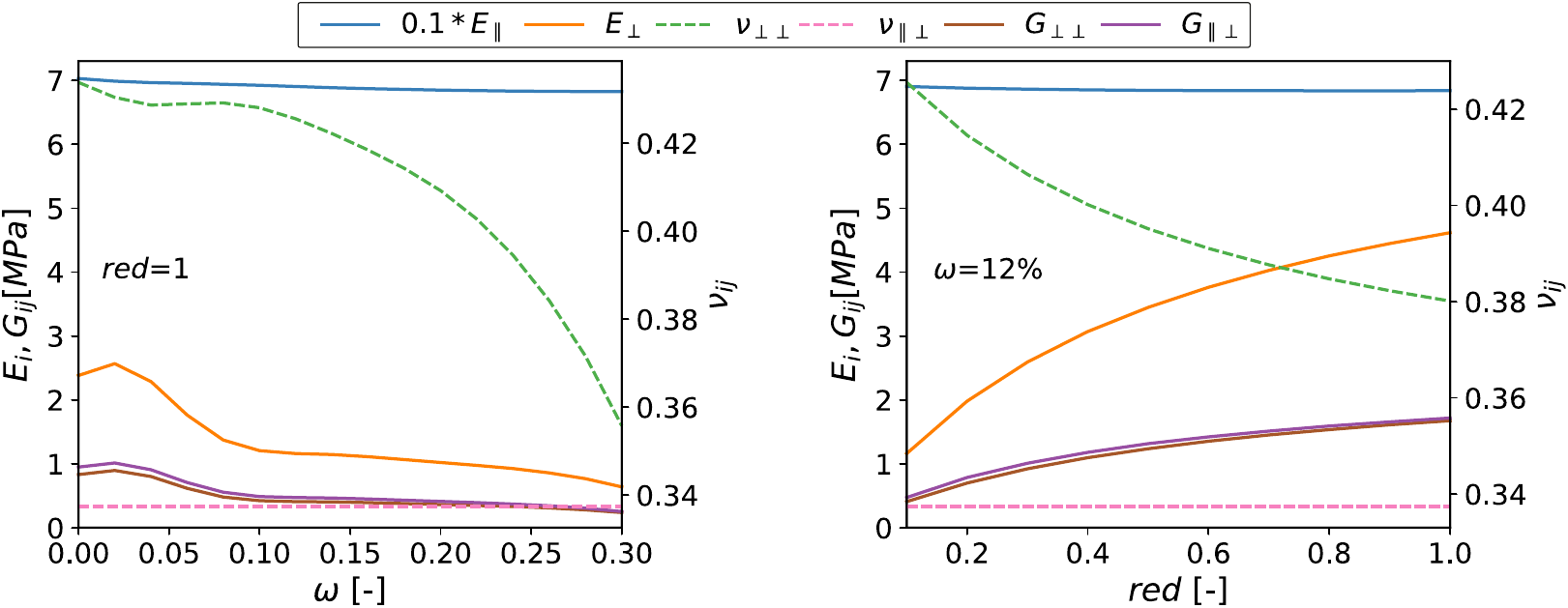}}
  \caption{\label{fig:1.5} Dependence of engineering constants on moisture $\omega$ and lignin degradation factor $red$ for the S2-layer in the layer coordinate system ($\parallel$ in micro fibril orientation, $\perp$ as the transverse).}
\end{figure}

To calculate the stiffness tensors of the different cell wall layers $\mathbf{C}^{CML},\mathbf{C}^{S1},\mathbf{C}^{S2},\mathbf{C}^{S3}$, the material stiffness $\mathbf{C}^{C},\mathbf{C}^{H}(u_b^H),\mathbf{C}^{L}(u_b^L)$ are combined with the respective volume fraction (see Tab.~\ref{tab:1}) and characteristics of their structural organization using a two-step mixing rule, since cellulose consists of both crystalline and amorphous regions, which are embedded within hemicelluloses to form microfibrillar aggregates \citep{Fernandes-etal-2011,Donaldson_2007}. To account for this, we determine the composite behavior of cellulose and hemicelluloses in the first step, which is then combined with the lignin in the second step. We consider the degrading effect of delignification by multiplying $\mathbf{C}^L$ with a reduction factor $red$. Then, the material properties of each layer are calculated with the Halpin-Tsai equations \citep{Halpin:1976} that can be expressed in short form as 
\begin{equation}
\label{eq:halpin}
P=P_{m} \frac{1+\zeta \eta v_{f}}{1- \eta v_{f}} \quad \text{with} \ \eta=\frac{p_{f}/p_{m}-1}{p_{f}/p_{m}+\zeta}, 
\end{equation}
where $P$ is a composite engineering constant (longitudinal Young's modulus $E_{\parallel}$, transverse Young's modulus $E_{\perp}$, longitudinal Poisson's ratio $\nu_{\parallel}$, longitudinal shear modulus $G_{\parallel}$, transverse $G_{\perp}$ shear modulus), $P_{f}$ and $P_{m}$ are the corresponding fiber and matrix constants, respectively, and $v_{f}$ is the fiber volume fraction. $\zeta$ is an empirical geometry parameter that depends on the reinforcement geometry and differs for composite properties ($\zeta \to \infty$ for $E_{\parallel}$ and $\nu_{\parallel}$, $\zeta=2$ for $E_{\perp}$, $\zeta=1$ for $G_{\parallel}$, $\zeta=\frac{K_{m}/G_{\perp,m}}{K_{m}/G_{\perp,m}+2}$ for $G_{\perp}$ with $K_{m}$ plane-strain bulk modulus of the matrix). Since the CML is assumed to be isotropic, its stiffness tensor is obtained by calculating the average of a series of rotated stiffness tensors around all axes. Note that due to the moisture dependence of the lignin and hemicelluloses stiffness, the layer stiffness is moisture dependent, as well. The resulting engineering constants from the elastic stiffness tensor in the layer coordinate system (see Fig.~\ref{fig:1new}) with dependence on moisture (for $red=1$) and on the reduction factor $red$ (for $\omega=12$) are given in Fig.~\ref{fig:1.5}. Given that, in this work, moisture states are not changing, calculations of hygro-expansion coefficients are not required.

For our model, we calculate an ESL stiffness $\mathbf{C}^{S}$ from the S1, S2, and S3 layers for each tracheid, excluding the CML since the tracheids are embedded into regions with $\mathbf{C}^{CML}$ properties. Since the stiffness tensor of S1, S2, and S3 layers is calculated within each layer coordinate system, it has to be rotated by the MFA around the $z^{cw}$-axis into the cell wall $x^{cw}=L-y^{cw}-z^{cw}$ coordinate system. For the combination of the different layer compliances in the cell-wall coordinate system, we strictly follow the description of the classical laminate theory to construct an ESL stiffness tensor of the secondary cell wall given in \cite{chou_etal_1972}. In this way, we can calculate and assign the material's linear elastic orthotropic stiffness of each tracheid, depending on its thickness. In reality, high deformation degrees, such as those in wood densification, can result in material non-linearity, driven by defilibration, fibril alignment, delamination, degradation, and multiple other phenomena that are neglected for now.
%%%%%%%%%%%%%%%%%%%%%%%%%%%%%%%%%%%%%%%%%%%%%%%%%%%%%%%%%%%%%%%%%%%%%%%%%%%%%%%%%%%%%%%%%%%%%%%%%%%%%%%%%%%%%%%%%%%%%%
\subsection{Tissue model generation}\label{sec:tissueGen}
\begin{figure*}[htb]
  \centering{\includegraphics[width=1\textwidth]{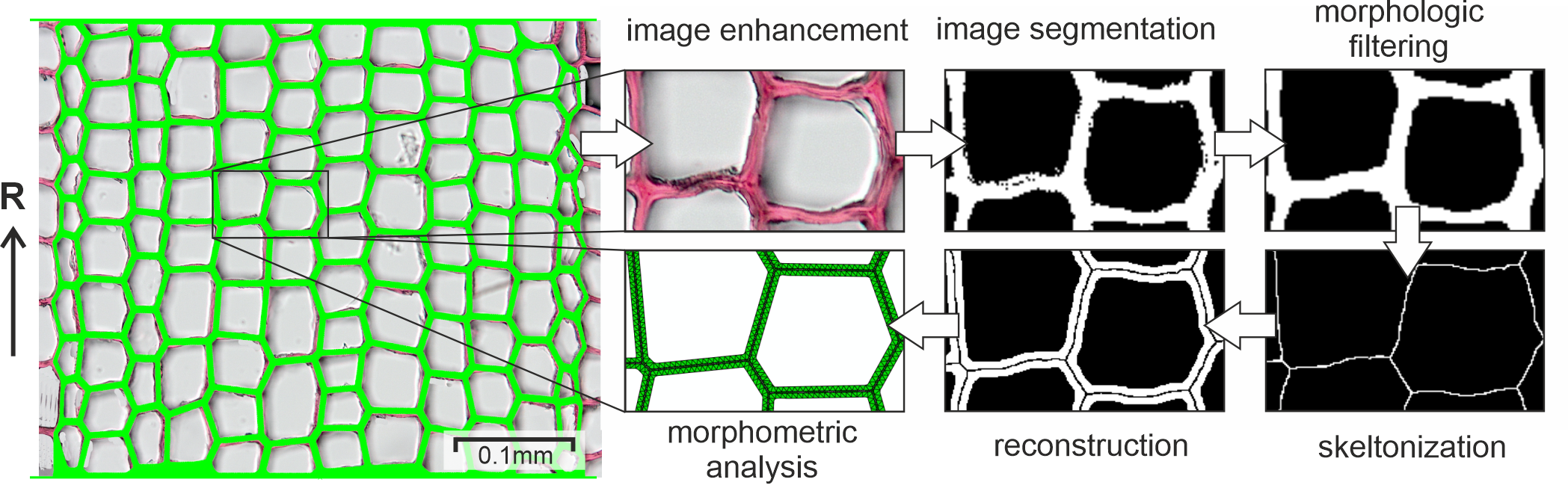}}
  \caption{\label{fig:2new} Processing of image information from the cross-sectional image to the FEM model.}
\end{figure*}
In cellular solids, the spatial material arrangement is at least equally important as the material behavior. Representative volumes (RVEs) can be obtained in various ways. One can determine morphometric distributions for the lumen ($L_T, L_R$) and cell wall ($l_T, l_R$) dimensions and then {\it in-silico} "grow" tissue sections by sampling from the distributions as demonstrated by \citet{mora_etal_2019}. The resulting tissue sections are representative in a statistical sense but miss details like tapered or rotated cells and many other features that cumulatively can affect compaction. Therefore, in this work, an image-based approach using the image processing toolbox of MATLAB is chosen. Note that the described approach is independent of the species.

Small 5x5\SI{}{\square\milli\metre} large sections of both native and compacted Norway spruce were polished using a rotary microtome (Leica RM2255) with an increment of \SI{30}{\micro\metre} per cut. The native sections were stained with safranin and transferred to microscope slides. Images were acquired with an Olympus BX51 microscope at 20x magnification, capturing the main anatomic features. When regions of interest exceeded the field of view, neighboring images were stitched. The compacted spruce samples with polished surfaces were sputtered with a \SI{15}{\nano\metre} Pt/Pd coating (Safematic, CCU-010) to guarantee conductance for Scanning Electron Microscopy (SEM). 

The process of the image analysis is shown in Fig.~\ref{fig:2new}. The intensity value pictures from the red channel of the microscopy images are first increased in resolution, and then contrast enhancement and histogram equilibration are applied to simplify the successive segmentation via Otsu's method. In the next step, morphological closing with a circular kernel is performed on the cell walls and after inversion of the picture on the lumen as well. Now, the image is ready to calculate the skeleton, which is subtracted from the image to segment individual cells. Defects from microtoming, rays, or impurities can disturb the automatic model generation significantly and need to be considered. Obvious defects were manually removed from the images before the analysis. Note that rays do not play a significant role in the densification of spruce and, therefore, can be ignored. As mentioned earlier, tracheid cell walls are assumed to be of uniform thickness. This assumption is not entirely correct, as $l_T \neq l_R$ \citet{mora_etal_2019,lanvermann_distribution_2013}. However, one can consider only one material per tracheid with discrete material orientation, defined by the lumen surface normal, as the consequences of this simplification are minor. The tissue is then reconstructed by "growing" the cell from the skeleton with uniform thickness. After all, the cellular topology, which is the most important feature with respect to mechanics, is captured correctly, as can be seen in Fig.~\ref{fig:2new}(left). In a final step, fibers that touch the boundary and spurious pixels from the skeletonization algorithm are removed. 
\begin{figure}[htb]
  \centering{\includegraphics[width=1\linewidth]{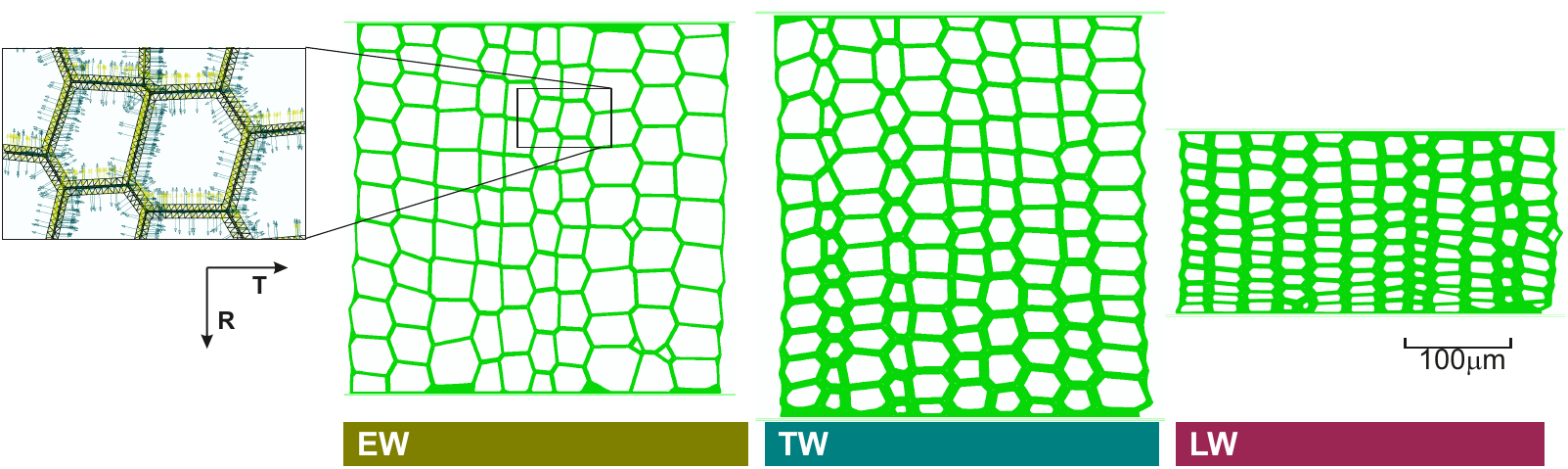}}
  \caption{\label{fig:3new} Models for the representative structures for early (EW), transition (TW), and late wood (LW). The magnification shows the mesh with CPE6M elements, and the local material coordinates with the orientation following the lumen wall.}
\end{figure}

All closed fibers are further analyzed, and the geometry is prepared to allow for an automated model generation in ABAQUS. For this purpose, the tracheid centroids are calculated and defined as mandatory reference points for the lumen fluid cavity condition for self-densification (see Sec.~\ref{sec:mat_num}). The average cell wall thickness is required to calculate the thickness of each S2 layer $t_{S2}$ by taking the ratio of the cell wall area to the length of the skeleton of the fiber. Only now, the cell wall stiffness $\mathbf{C}^{S}$ of each tracheid can be calculated. Cubic spline interpolations of the perimeter pixels give the outer and inner perimeters of the cell. The outer perimeter of each tracheid is down-scaled by the CML thickness $t_{CML}$ so the CML matrix can fill the space between tracheids. The inner perimeter (or lumen perimeter) is later on required to assign the discrete material orientation of the orthotropic material and as a surface for the contact calculation to avoid inter-penetration in compaction.
\begin{figure*}[htb]
    \centering
    \includegraphics[width=1\linewidth]{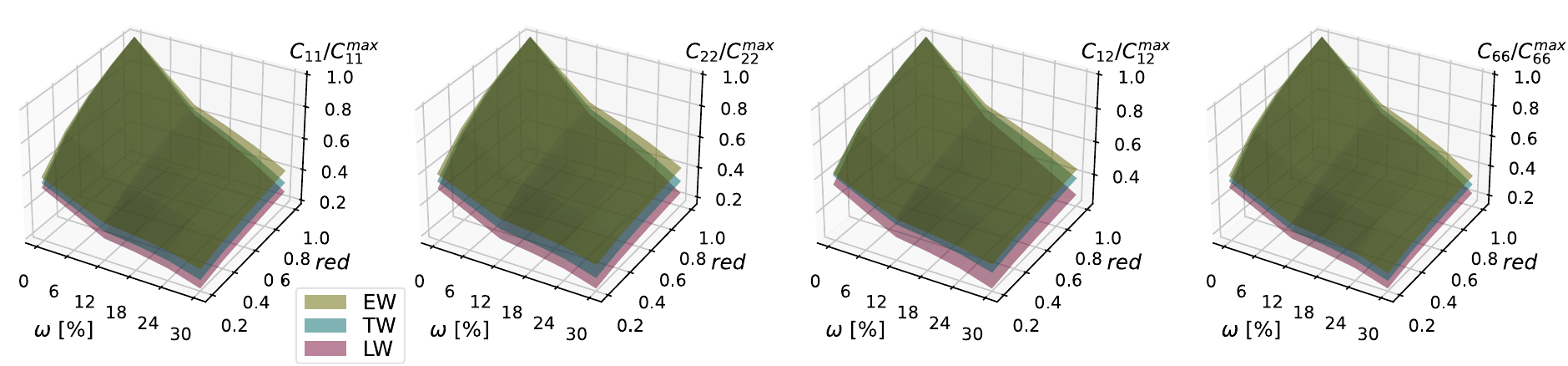}
    \caption{Dependence of the coefficients of the stiffness tensor $C_{ij}$ for EW, TW, and LW on moisture and lignin degradation normalized with the initial coefficients in Tab.~\ref{tab:4}. The stiffness coefficients are expressed in the tissue coordinate system (1-direction = tangential (T), 2-direction = radial (R)).}
    \label{fig:1.6}
\end{figure*}

The model generation is automatized using Python scripting. Initially, each tracheid is generated from the closed splines of the inner and outer perimeter. All tracheids are then merged into a single part, while each represents an individual section for material assignment with the orthotropic stiffness tensor from the ESL calculation. As the isotropic CML requires a distinct material section, it is generated as a separate part and connected via tie connections to the merged tracheids. Examples of final models of LW, TW, and EW are shown in Fig.~\ref{fig:3new}. All lumen surfaces have self-contact conditions that are assumed to be frictionless for tangential relative motion and as hard contact perpendicular to the surfaces, avoiding cell wall interpenetration. A general contact is defined for the outer perimeter of the matrix to avoid model overlaps for large compaction degrees. In principle, one faces a generalized plane strain case,  however, we keep the model formulation as plane strain with unit thickness due to the strong stiffness contrast to the longitudinal direction. This way, 6-node triangular explicit elements for plane strain (CPE6M) can be used for the mesh generation. 

As all simulations are plain strain, the stiffness tensor is reduced to 
\begin{equation}
\mathbf{C}=\begin{bmatrix}
C_{11} & C_{12} & 0 \\ 
  & C_{22} &  0 \\ 
sym. &      & C_{66}
\end{bmatrix}.
\end{equation}
in the system coordinate system, where 1 corresponds to the T-direction, 2 to the R-direction, and accordingly, $C_{66}$ represents the shear in the RT-plane. 

First, the initial stiffness of the different tissue types and their relative dependence on moisture and lignin degradation are calculated. In Tab.~\ref{tab:4}, the dry, non-degraded reference values and engineering constants are given. Those are used to normalize the values in Fig.~\ref{fig:1.6}. The sensitivity of LW to moisture and lignin degradation is highest for all components, with the dependence being higher on moisture than lignin degradation. Interestingly, the shear component shows the least difference between the tissue types. After characterizing the small-strain behavior, we can numerically densify the models.

\begin{table}[htb]
\centering
\caption{Coefficients of the stiffness tensor in a fully dry and engineering constants at $\omega=24$ both in the non-degraded state in GPa. The values are expressed in the tissue coordinate system (1-direction = tangential (T), 2-direction = radial (R)).}
\label{tab:4}
\begin{tabular}{l c c c c c c c}
\hline
  & $C_{11}$ & $C_{22}$ & $C_{12}$ & $C_{66}$ & $E_{T}$ & $E_{R}$ & $G_{RT}$\\
\hline\hline
EW & 0.862 & 0.899& 0.339 & 0.016 &0.423 & 0.447& 0.01 \\ 
TW & 2.239 & 1.626& 0.665 & 0.083 & 1.153& 0.795& 0.051 \\
LW & 4.336 & 3.429 & 1.207 & 0.428 & 2.23&1.647 &0.25 \\ \hline
\end{tabular}
\end{table}
%%%%%%%%%%%%%%%%%%%%%%%%%%%%%%%%%%%%%%%%%%%%%%%%%%%%%%%%%%%%%%%%%%%%%%%%%%%%%%%%%%%%%%%%%%%%%%%%%%%
\subsection{Densification Simulations} \label{sec:mat_num}
All simulations are performed as dynamic explicit, using mass scaling to increase time increments. Since inertial effects are negligible, one refrains from adopting the density to different moisture states and delignification levels and uses a density of \SI{1500}{\kilogram\per\cubic\metre}. All simulations are performed on the representative sections (see Fig.~\ref{fig:3new}), but the procedure also works on other micro-structural images in general. We always densify in the radial direction, as this is the technologically relevant case. Additionally, the effect of delignification and moisture is considered (see Fig.~\ref{tab:3}. Note that delignification would modify sorptivity and the partition of bound water between hemicelluloses and lignin ($u_b^L,u_b^H$, Eqs.~\ref{eq:8}). However, we stick to the unmodified case for clarity since those details have not been studied so far. In addition, the stiffness reduction of lignin must not necessarily be caused by delignification but could also be achieved by heat, as the glass transition temperature of lignin is around \SI{70}{\celsius} \cite{engelund_critical_2013}.
\begin{figure}[htb]
  \centering{\includegraphics[width=1\linewidth]{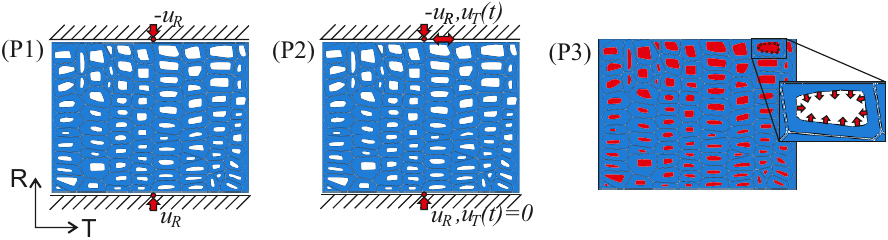}}
  \caption{\label{fig:4new} Boundary conditions for the protocols (P1-P3) with control points.}
\end{figure}

As the focus of this work is on alternative densification protocols and procedures, we define the following protocols, shown in Fig.~\ref{fig:4new}:
\begin{itemize}
    \item[(P1):] \underline{Mechanical densification}: This classical densification is realized by assigning an opposite vertical displacement control on reference points that are kinematically coupled to top and bottom plates, contacting the model. Vacuum densification can be obtained from these simulations as well, except that pressure would correspond to the average reaction force. 
    \item[(P2):] \underline{Transverse vibration-assisted mechanical densification}: Identical to (P1) but with additional horizontal kinematic coupling of reference points and harmonic horizontal oscillatory excitation $u_T=amp\cdot\sin(2\pi\cdot \mathit{freq} \cdot t)$ at variable amplitude $amp$ and frequency $\mathit{freq}$.  
    \item[(P3):] \underline{Self-densification}: All external surfaces are free to move, but shrinkage of lumen-filling fluid can collapse tracheids. In the simulations, this is realized with fluid cavity conditions with compressible fluid (bulk modulus $E_{fl}$=\SI{10e-6}{\giga\pascal} and density $\rho_{fl}$= \SI{500}{\kilogram\per\cubic\metre}). Its contraction is controlled by temperature $T=[0,15]$\SI{}{\celsius} for EW, $T=[0,35]$\SI{}{\celsius} for TW, and $T=[0,85]$\SI{}{\celsius} for LW, with linear heat expansion $\alpha_T$=\SI{-1000}{\per\celsius}.
\end{itemize}

\begin{table*}[htb]
\caption{\label{tab:3} Summary of simulation for protocols (P1-P3) with ranges of control parameters.}
\centering
\begin{tabular}{lllllll}
\hline
    & $\omega$ [\%] & $red$ & $u_R$ & $u_T$ & $\mathit{freq}$ & $T$\\\hline\hline
    P1 & 0,6,12,18,24,30 & 0.2,0.4,0.6,0.8,1 & $u_R$ & & & \\
    P2 & 24 & 0.8 & $u_R$ & $amp$ & $\mathit{freq}$ & \\
    P3 & 30 & 0.2 & & & & 15, 35, 85 \\\hline
\end{tabular}
\end{table*}

A summary of simulation cases is given in Fig.~\ref{fig:4new} and Tab.~\ref{tab:3}. The simulations are juxtaposed by comparing the evolution of porosity, work, compaction force, or elastic energy, as well as qualitatively by characteristic emerging folding patterns.

%%%%%%%%% --- RESULTS AND DISCUSSION ---
\section{Results and discussion}
Densification, namely the sequential collapse of cellular structures, is initiated by micro-structural instabilities. These processes are not apparent at the larger scale of observation when the complex interactions are hidden in the force-compaction relation. To study the self-organized folding pattern formation in different tissue types as well as ways to manipulate them, large enough representative models with a sufficient number of tracheids and ways to solve the systems dynamically, including geometric non-linearity from large deformations and displacements, are prerequisites. First, we show the explicit dynamic simulation results of radial densification for different tissue types, moisture, and delignification states. After a phenomenological characterization of the evolution, we look at the morphogenetic differences observed in alternative densification procedures like shear-assisted or self-densification.
%%%%%%%%%%%
\subsection{Mechanical densification of spruce tissues} \label{sec:res_exp}
\begin{figure}[htb]
  \centering{\includegraphics[width=1\textwidth]{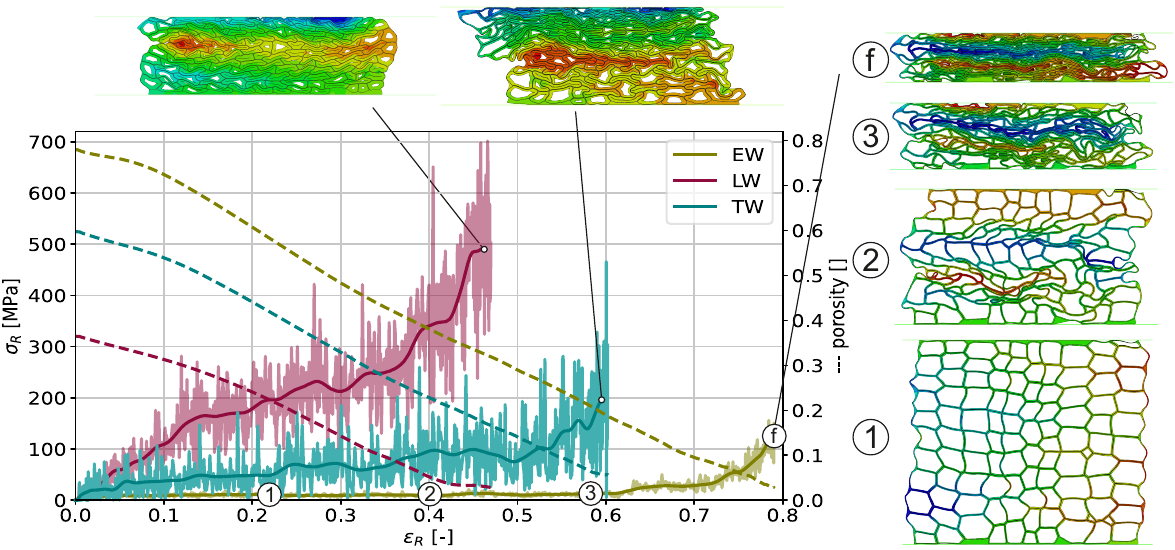}}
  \caption{\label{fig:dens_simple} Densification evolution and morphogenesis of folding patterns in early wood for $\omega=$\SI{24}{\percent} and no delignification. Colors indicate horizontal displacements for increased visibility.}
\end{figure}

\begin{figure}[htb]
  \centering{\includegraphics[width=0.8\textwidth]{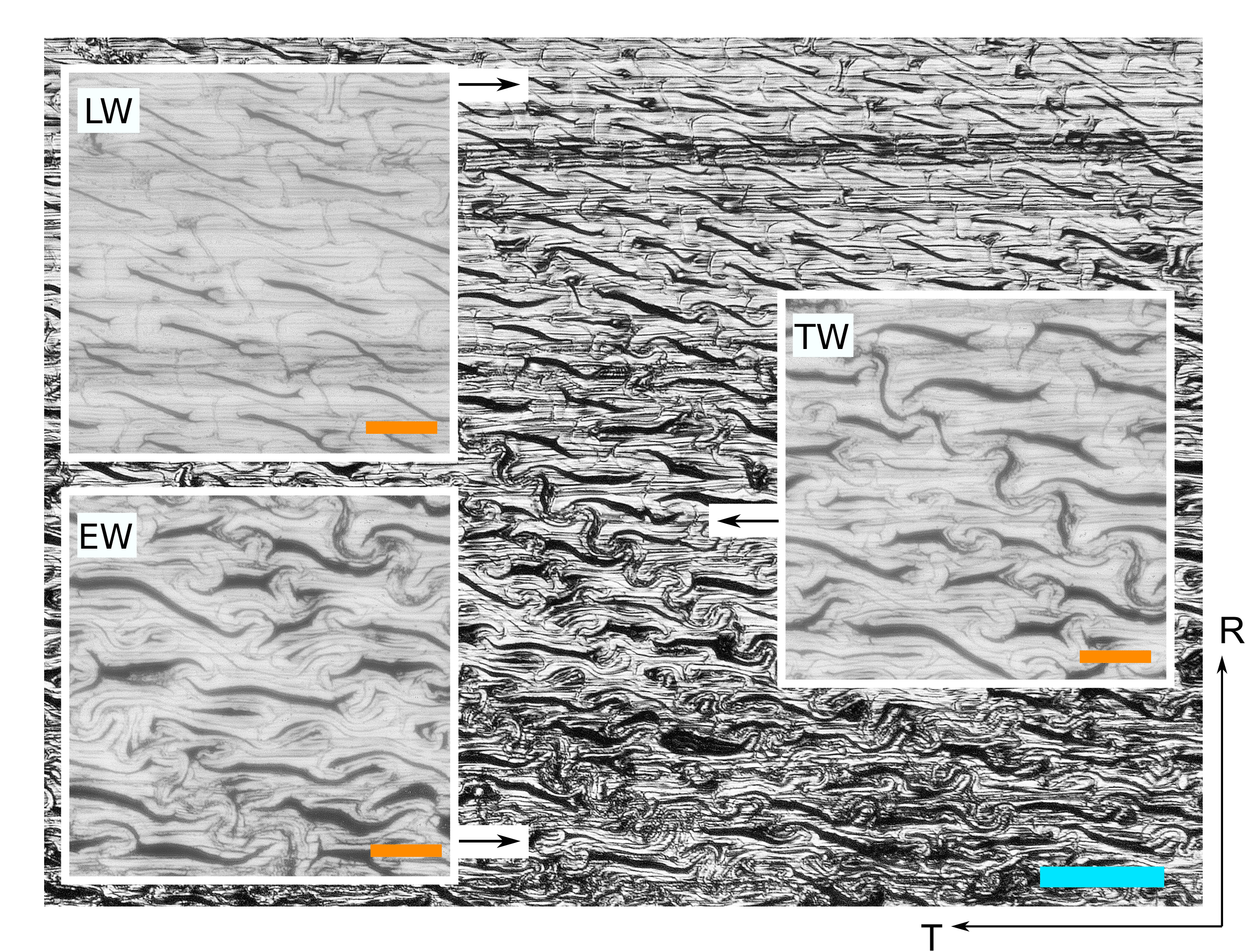}}
  \caption{\label{fig:microscopy} Mechanically densified Norway spruce tissues across a growth ring. The scale bar in the overview image corresponds to \SI{50}{\micro\meter} (cyan), while the ones in the magnifications relate to \SI{20}{\micro\meter} (orange).}
\end{figure}

The simple radial mechanical densification is at the core of most densification processes, as well as of relevance for certain dowel connections in timber structures under high loads (\cite{sotayo_review_2020}). The process for EW, TW, and LW sections is simulated at different moisture contents $\omega$ and delignification degrees represented by $red$. Since the simulations are displacement controlled, the monitored properties are the reaction force on the boundaries that are used to calculate the engineering stress $\sigma_R$ for a given radial strain $\varepsilon_R$, as well as the porosity calculated from the lumen area normalized by the initial total area of the representative section. For $\omega=$\SI{24}{\percent} and no delignification ($red=0$), the process evolution is given in Fig.~\ref{fig:dens_simple} along with snapshots of the related morphological observations for EW. It is particularly striking to observe that both EW and TW exhibit a distinct compaction plateau that is missing for LW. This plateau forms due to a state of equilibrium between hardening and softening mechanisms, until the contacting cell walls of collapsed tracheids increasingly form direct load paths. A pronounced hardening behavior and a strong increase in the tangent stiffness mark this transition. For LW, the low porosity and elongated lumen shapes, along with the thick cell walls, prohibit such a collapse-driven plateau formation. When increasing the load (Fig.~\ref{fig:dens_simple}), LW still deforms in a linear fashion when EW and TW already exhibit massive densification. It comes as no surprise that the majority of the densification process is predominantly driven by the compaction of early wood (EW) and transition wood (TW) tissue, characterized by their relatively large lumen and thin cell walls, which agrees with the densified structures observed under the microscope shown in Fig.~\ref{fig:microscopy}. These structural features make EW and TW more susceptible to densification, as their cellular architecture allows for significant deformation under applied loads compared to denser, thicker-walled tissues. Note that the final tangent stiffness is similar for all three tissue types. However, simulations at the final densification states require extremely small time increments to succeed and are often accompanied by numerical instabilities due to highly distorted meshes. 

The simulations provide technologically significant insights into the morphogenetics of folding pattern formation. Typically, radial cell walls are strongly aligned as a direct consequence of their origin from the same mother cell located in the cambium. This structural alignment leads to relatively straight, though slightly undulating, load paths for radial force. Buckling is a geometric instability where slender elements suddenly transmit much less force. As cell walls of \textbf{early wood} are thin, buckling loads are reached at an early stage of loading throughout the cellular structure. However, this first phase is characterized by spatially weakly correlated radial cell wall buckling, as in the vertical direction, the load is taken off the force chain, and in the horizontal direction, thin walls are inefficient in transmitting moments. As a consequence, dispersed isolated initial buckling of radial walls is observed (see Fig.~\ref{fig:dens_simple}.1). This picture changes radically for further densification. In the cellular structure, load redistribution to neighboring force lines occurs. Additionally, the tangential displacement from the buckled cell wall row is transmitted via the tangential walls directly to the adjacent cell wall row. This motion reduces the critical load in those adjacent regions, triggering - in combination with the enhanced load from the redistribution - a correlated buckling along the tangential direction and consequently the formation of a localized compaction band (Fig.~\ref{fig:dens_simple}.2). The intricate interplay of forces and instabilities during the densification process continues, particularly as the rotation of cell corners — where three or four cells come into contact — introduces additional complexity. The rotational movement at these junctions exerts a counter-moment on the radially adjacent row of cells, collaboratively inducing buckling in the opposite direction. This reciprocal buckling behavior, driven by the interconnected structure, reinforces the propagation of deformation across the cellular network, further resulting in regularity in the densification patterns and adding to the complexity of the overall mechanical response (Fig.~\ref{fig:dens_simple}). With increased densification, this sequential compaction by densification in additional compaction bands continues (Fig.~\ref{fig:dens_simple}.3) until full compaction is reached throughout the material by direct force transmission via contact of cell walls in their lumen (Fig.~\ref{fig:dens_simple}.f). In \textbf{transition wood}, the density gradient plays an additional role, where densification starts from the lower density zones, as described for EW. In \textbf{late wood}, one cannot observe sequential densification in compaction in bands. Here, one observes a pretty uniform densification over the entire volume of the system, as cell walls are rather bulky.

%Effect of moisture and delignification
\begin{figure}[tb]
  \centering{\includegraphics[width=0.7\textwidth]{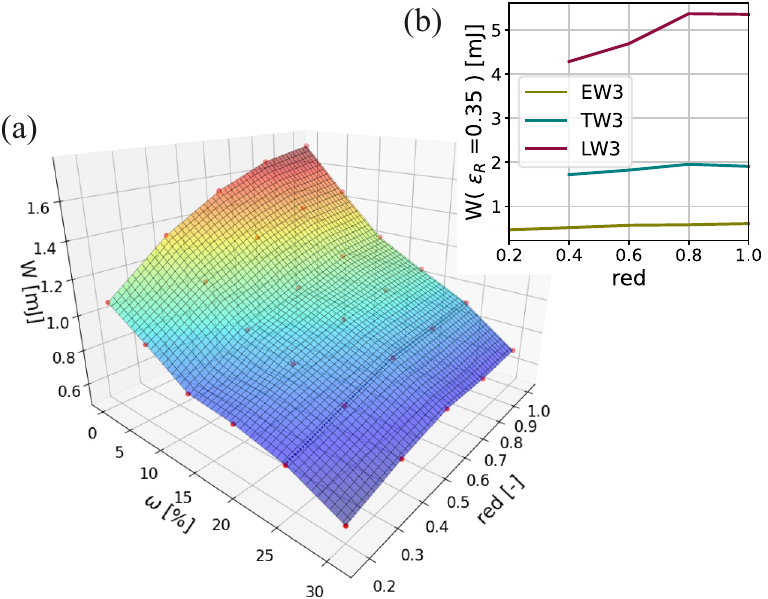}}
  \caption{\label{fig:dens_work} (a) Total work of densification at \SI{50}{\percent} vertical strain for EW and at (b) \SI{35}{\percent} vertical strain for the different tissues at $\omega=$\SI{24}{\percent}.}
\end{figure}
When comparing densification across different moisture contents and degrees of delignification, the systems exhibit remarkably similar morphogenetic behavior. However, the differences become apparent when calculating the work $W$ required for densification up to a given vertical strain for comparability, shown in Fig.~\ref{fig:dens_work}. As one expects, the work $W$ needed for LW far exceeds that of TW, while EW requires the least effort (Fig.~\ref{fig:dens_work}(b)) \cite{laine_wood_2016}. Notably, for all tissue types, the reduction in required work due to increased moisture content surpasses that caused by delignification (shown for EW in Fig.~\ref{fig:dens_work}(a)).

Densification is often not achieved through displacement control but by applying a defined pressure. One of the simplest methods works with a press or a vacuum pump to depressurize the sealed wood piece, allowing densification of delignified wood to occur solely due to atmospheric pressure. However, atmospheric pressure alone may not always be sufficient to trigger densification fully, and in some cases, only partial densification can be achieved. The simulation results for displacement-controlled compaction, exemplified in Fig.~\ref{fig:dens_simple}, provide the information between applied pressure and densification degree. From this information, one can extract the densification profiles across growth rings as a function of the uniform pressure for any given tissue distribution in growth rings. It is evident from Fig.~\ref{fig:dens_simple} that the stiff late wood can withstand atmospheric pressure without significant densification, highlighting its resistance compared to other tissue types that would collapse first. As localized sequential densification bands progressively form, partial densification will result in distinct density bands in EW and TW.
%%%%%%%%%%%%%%%%%%%%
\subsection{Transverse vibration-assisted mechanical densification}
An intricate and complex interplay of instabilities and load redistribution can be observed in the initial phase of densification. As tracheid walls begin to deform, localized instabilities emerge, causing the load to be redistributed dynamically across neighboring structures. This process initiates cascades of interactions, with buckling, bending, and shifting forces propagating through the material, laying the groundwork for the emerging densification bands. One can try to influence these processes by introducing additional transverse oscillations on the loading plates during densification. The idea would be that oscillatory motion promotes more uniformity in the buckling morphologies by reducing spatial correlations in buckling. Ideally, densification zones would not form, and one could expect a more homogeneous densification profile throughout the compacted systems. Consequently, the impact on buckling instabilities would be most visible in early wood by reducing the first phase densification work and allowing for a homogeneous packing density. 
% Figure with work
\begin{figure}[htb]
  \centering{\includegraphics[width=0.8\textwidth]{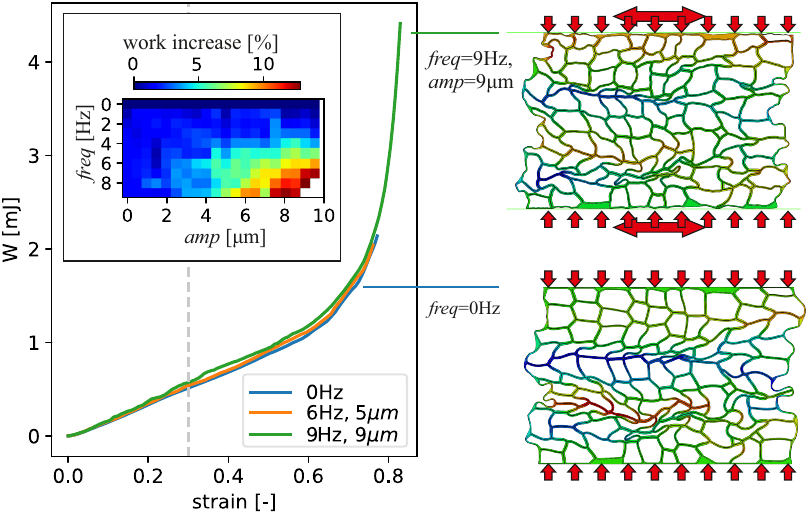}}
  \caption{\label{fig:samd} Work of compaction with shear-vibration assisted densification for $\omega=$\SI{24}{\percent} and $red$=\SI{0.8}{}, work increase in \% with respect to simple densification for densification at \SI{30}{\percent} strain with variable frequency $f$ and amplitude $amp$, as well as two snapshots during early state densifitcation with and without transvese vibration assistance. Colors represent horizontal displacements for a clearer visualization.}
\end{figure}

Since different frequencies and amplitudes can produce varying effects, we conducted simulations across a frequency range of $\mathit{freq}$=\numrange{0}{9}\SI{}{\hertz} and amplitudes between $amp=\numrange{0}{9}$\SI{}{\micro\metre}, which corresponds to approximately \SI{35}{\percent} of the tangential tracheid width. This range was selected to investigate the impact of oscillatory parameters on the densification process and to determine the conditions that most effectively alter folding morphogenesis. In Fig.~\ref{fig:samd}, the dependence of the compaction work on oscillation is given. Interestingly, the work for compaction up to a given strain is always larger with transverse vibrations than without. What sounds surprising is somewhat logical, as localization in shear bands is delayed, and lattice vibrations would result in system enlargement that compaction hinders. Note that missing values due to the abortion of simulations were interpolated from the results of successful runs. Also, inertial effects might be exaggerated due to mass scaling.

The differences are not striking when comparing morphologies at identical compaction states (see Fig.~\ref{fig:samd}). In shear-assisted densification, the entire structure is in motion, delaying the localization of bands by reducing spatial correlations. However, similar densification patterns form as in the simple densification. When comparing the tracheids near the load introduction plates on the top and bottom, a clear difference is evident, with many more cell walls being buckled due to transverse vibration that still appear unbuckled in the simple densification.
%%%%%%%%%%%%%%%%
\subsection{Self-densification}
For self-densification to occur, the lumen must be filled with an incompressible, extensively shrinking fluid. In practice, this can be achieved by infiltrating a hydrogel, such as gelatin (\cite{koch-etal-2024}). Upon drying, the volumetric shrinkage of the hydrogel directly translates into the densification of the wood tissue. To date, this process has only been successful with delignified wood, and the hydrogel must fully infiltrate both the lumen and the delignified cell walls to prevent debonding of the drying gelatin from the cell wall surface. One significant advantage of this self-densification process is the ability to mold the material into any desired shape in its wet state. Once dried, the gelatin locks the material, allowing it to self-densify without additional dyes or finishing treatments. Consequently, one must utilize the cell wall properties of the water-saturated state ($\omega$=\SI{30}{\percent}) with the lowest lignin content feasible in simulations ($red=$0.2). Comparison with tensile experiments on water-saturated cellulose scaffolds from delignified Norway spruce \cite{frey_etal_2019} give values of about $E_L=$\SI{54.5}{\mega\pascal}, while the bulk spruce is in the order of \SI{10}{\giga\pascal} at $\omega$=\SI{12}{\percent}. From this ratio, we calculate the scaling factor for the compliance tensor of the tissue models to obtain an estimate of the cavity pressure of the shrinking gel. Interestingly, one obtains quite unrealistic cavity pressures of up to \SI{0.8}{\mega\pascal} below the atmospheric pressure, which would be way below the vapor pressure of the water-based gelatine. However, if one considers the extremely low strength values in the longitudinal direction, determined by \cite{frey_etal_2019} as \SI{125}{\kilo\pascal}, it is evident that the cell walls quickly lose rigidity, already at the beginning of the densification. This explains the low values of the cavity pressure, which assumes intact cell walls for the entire process of densification without any loss of stiffness. Nevertheless, the morphogenesis should be rather material-independent, as well as the relative porosity evolution for the different tissue types.
\begin{figure*}[htbp]
  \centering{\includegraphics[width=1\textwidth]{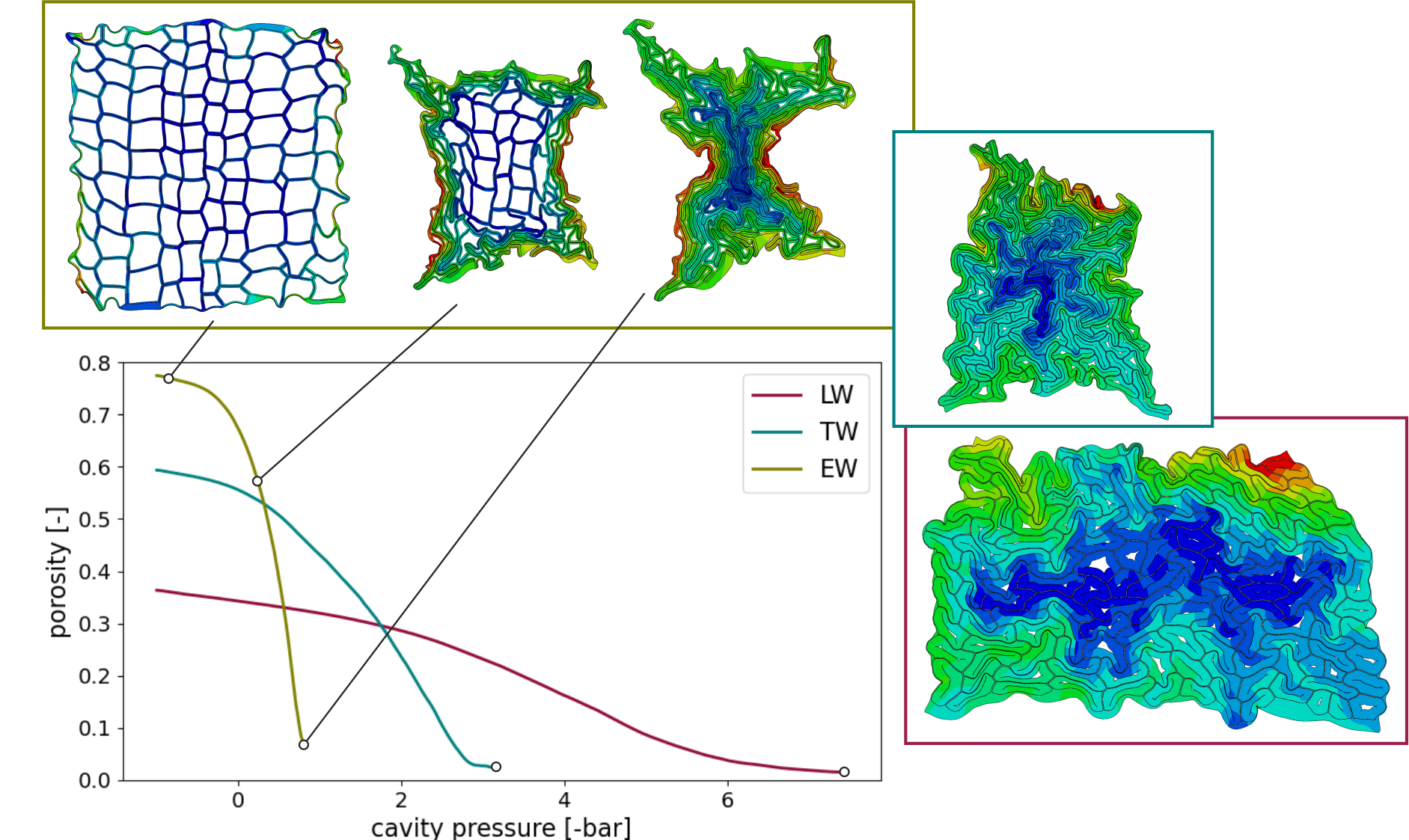}}
  \caption{\label{fig:selfdens} Emerging morphologies in self-densification for EW, TW, and LW. Colors represent displacement magnitudes to enhance visibility.}
\end{figure*} 

In the absence of external constraints, the folding pattern evolves through undirected cell collapse, progressing from the outer surfaces toward the interior, which must be explained by the fact that outer cells have no counter-pressure from the surface and, therefore, collapse first (see Fig.~\ref{fig:selfdens}). Depending on the available pressure, the compaction can be incomplete, forming a more porous zone near the center, especially if the outer densified regions establish a stable surface. This stability is often driven by arching effects, where the outer layers bear the load and resist further deformation, leaving the core less compacted but exhibiting a compacted surface. In simulations, the internal pressure can drop below the vapor pressure, particularly within thick-walled late wood tracheids with a small, circular lumen. This must be expected since composite mixing rules, described in Sec.~\ref{sec:cellWall}, result in too low compliances for extreme mechanical contrasts between constituents. However, this has little effect on the qualitative collapse behavior for self-densification. When comparing the densification states of different tissue types for a given pressure (see Fig.~\ref{fig:selfdens}), it is evident that the EW fully densifies first, followed by the TW and LW. For finite pressure values, one must expect higher porosity in late and transition wood.

% --- CONCLUSION ---
\section{Conclusions}
The densification of cellular solids, such as spruce wood, is characterized by a rich array of micro-mechanisms involving a complex interplay of instabilities and forces across different tissue types. Dynamic, explicit simulations are essential to simulate this process accurately. These simulations must capture the moisture-dependent behavior of tracheid walls, ranging from early to late wood, while incorporating the mechanically relevant characteristics of the wood's microstructure, including its inherent disorder. A comprehensive approach that starts from real micro-structures is crucial to replicating the intricate interactions that drive densification in such materials. From a technological perspective, mechanical densification in the radial direction is the most relevant, as it aligns with the sequential arrangement of growth rings, beginning with early wood, followed by transition wood, and late wood. Since the densification behavior of each tissue type is mainly uncoupled from one another, they can be simulated individually, allowing for a more detailed analysis of their respective densification mechanisms. Ultimately, the overall densification behavior can be estimated by considering the relative contributions of each tissue type to the growth rings, enabling a more accurate prediction of the material’s response under applied loads.

We showed that the compaction behavior is dominated by sequential buckling collapse with transverse propagation via cellular mechanics. Utilizing real micro-structures with randomness adds to the realism of the simulations. In our simulations, we assumed the composite cell wall material to be linear elastic without degradation. Consequently, all observed non-linearity must originate from geometric non-linearity, like buckling or contacts. However, one has to be aware that large cell wall deformations would result in intra-wall damage, increasing the compliance of the cell walls depending on the loading history. This is still an open field of research. The consequences of this assumption are visible in transverse vibration-assisted mechanical densification, where vibrations could loosen the cell walls to reduce the work of densification. Also, for self-densification, intra-wall stresses exceed the strength of the delignified, swollen cellular scaffold, resulting in unrealistically high cavity pressure for densification. Nevertheless, our simulations revealed morphogenetic mechanisms and resulting morphological features that increase the understanding of the potential and limits of densification. We showed that even though densification appears to be a self-organizing process at the tissue scale, we can influence the morphogenesis of folding pattern formation in various ways. The ability to tailor the properties of densified wood through modifications and targeted densification enables densified wood products to meet a wide range of performance requirements, making it a highly versatile and sustainable option for numerous industries. Through our analysis and simulations, we enrich the understanding of the mechanisms driving densification, paving the way for more controlled and efficient applications of this technique in industrial settings.

\backmatter
%%%%%%%%%%%%%%%%
\bmhead{Acknowledgements}
The financial support from the Swiss National Science Foundation under SNF grant 200021\_192186 "Creep behavior of wood on multiple scales" is acknowledged, as well as the preliminary work by Felix Engel.

%\bibliographystyle{bst/sn-basic.bst}
%\bibliography{references}

\end{document}